\PassOptionsToPackage{unicode}{hyperref}
\PassOptionsToPackage{hyphens}{url}
\documentclass[
]{article}
\usepackage{amsmath,amssymb}
\usepackage{iftex}
\ifPDFTeX
  \usepackage[T1]{fontenc}
  \usepackage[utf8]{inputenc}
  \usepackage{textcomp} 
\else 
  \usepackage{unicode-math} 
  \defaultfontfeatures{Scale=MatchLowercase}
  \defaultfontfeatures[\rmfamily]{Ligatures=TeX,Scale=1}
\fi
\usepackage{lmodern}
\ifPDFTeX\else
\fi
\IfFileExists{upquote.sty}{\usepackage{upquote}}{}
\IfFileExists{microtype.sty}{
  \usepackage[]{microtype}
  \UseMicrotypeSet[protrusion]{basicmath} 
}{}
\makeatletter
\@ifundefined{KOMAClassName}{
  \IfFileExists{parskip.sty}{%
    \usepackage{parskip}
  }{
    \setlength{\parindent}{0pt}
    \setlength{\parskip}{6pt plus 2pt minus 1pt}}
}{
  \KOMAoptions{parskip=half}}
\makeatother
\usepackage{xcolor}
\usepackage[margin=1in]{geometry}
\usepackage{color}
\usepackage{fancyvrb}

\DefineVerbatimEnvironment{Highlighting}{Verbatim}{commandchars=\\\{\}}
\usepackage{framed}
\definecolor{shadecolor}{RGB}{248,248,248}
\newenvironment{Shaded}{\begin{snugshade}}{\end{snugshade}}

\newcommand{\AttributeTok}[1]{\textcolor[rgb]{0.13,0.29,0.53}{#1}}

\newcommand{\ConstantTok}[1]{\textcolor[rgb]{0.56,0.35,0.01}{#1}}

\newcommand{\DecValTok}[1]{\textcolor[rgb]{0.00,0.00,0.81}{#1}}

\newcommand{\FunctionTok}[1]{\textcolor[rgb]{0.13,0.29,0.53}{\textbf{#1}}}

\newcommand{\NormalTok}[1]{#1}

\newcommand{\OtherTok}[1]{\textcolor[rgb]{0.56,0.35,0.01}{#1}}

\newcommand{\SpecialCharTok}[1]{\textcolor[rgb]{0.81,0.36,0.00}{\textbf{#1}}}

\newcommand{\StringTok}[1]{\textcolor[rgb]{0.31,0.60,0.02}{#1}}

\usepackage{graphicx}
\makeatletter
\newsavebox\pandoc@box
\newcommand*\pandocbounded[1]{
  \sbox\pandoc@box{#1}%
  \Gscale@div\@tempa{\textheight}{\dimexpr\ht\pandoc@box+\dp\pandoc@box\relax}%
  \Gscale@div\@tempb{\linewidth}{\wd\pandoc@box}%
  \ifdim\@tempb\p@<\@tempa\p@\let\@tempa\@tempb\fi
  \ifdim\@tempa\p@<\p@\scalebox{\@tempa}{\usebox\pandoc@box}%
  \else\usebox{\pandoc@box}%
  \fi%
}
\def\fps@figure{htbp}
\makeatother
\NewDocumentCommand\citeproctext{}{}

\makeatletter
 \let\@cite@ofmt\@firstofone
 \def\@biblabel#1{}
 \def\@cite#1#2{{#1\if@tempswa , #2\fi}}
\makeatother
\newlength{\cslhangindent}
\newlength{\csllabelwidth}
\newenvironment{CSLReferences}[2] 
 {\begin{list}{}{%
  \setlength{\itemindent}{0pt}
  \setlength{\leftmargin}{0pt}
  \setlength{\parsep}{0pt}
  \ifodd #1
   \setlength{\leftmargin}{\cslhangindent}
   \setlength{\itemindent}{-1\cslhangindent}
  \fi
  \setlength{\itemsep}{#2\baselineskip}}}
 {\end{list}}
\usepackage{calc}

\usepackage{bm}
\usepackage{pdfpages}
\usepackage{hyperref}
\usepackage{xcolor}
\usepackage{booktabs}
\usepackage{multicol}
\newcommand{\proglang}[1]{\textbf{\texttt{#1}}}
\newcommand{\pkg}[1]{\textbf{#1}}

\hypersetup{colorlinks=true, citecolor=blue, linkcolor=blue, urlcolor=magenta}
\usepackage{bookmark}
\IfFileExists{xurl.sty}{\usepackage{xurl}}{} 
\hypersetup{
  pdftitle={A joint model of correlated ordinal and continuous variables},
  pdfauthor={Laura Vana and Rainer Hirk},
  hidelinks,
  pdfcreator={LaTeX via pandoc}}

\title{A joint model of correlated ordinal and continuous variables}
\usepackage{etoolbox}
\makeatletter
\providecommand{\subtitle}[1]{
  \apptocmd{\@title}{\par {\large #1 \par}}{}{}
}
\makeatother
\subtitle{This version: 2024-11-05}
\author{Laura Vana and Rainer Hirk}
\date{}

\begin{document}
\maketitle

\section{Introduction}\label{introduction}

The analysis of mixed type responses is a relevant topic in various
fields of research where one often jointly collects various continuous,
binary or ordinal outcomes possibly with some observations missing.

The joint modeling of outcomes is sometimes preferred over a separate
analysis of the different responses given that the association between
different outcomes can be captured is capture by the joint model
specification. The advantage of this approach is that answers to various
research questions can be obtained in one go, thus eliminating the need
for multi-step procedures to combine results from various separate
analyses.

Various approaches have been proposed for building such joint models.
The random effects approach relies on specifying correlated random
effects among the different outcomes (Ivanova, Molenberghs, and Verbeke
2016). Conditional on the random effects, the responses are assumed to
be independent. Another popular approach is assuming a multivariate
distribution directly on the errors of the model corresponding to the
different responses.

In this paper we build a joint model which can accommodate for binary,
ordinal and continuous responses, by assuming that the errors of the
continuous variables and the errors underlying the ordinal and binary
outcomes follow a multivariate normal distribution. We employ composite
likelihood methods to estimate the model parameters and use composite
likelihood inference for model comparison and uncertainty
quantification. The complimentary R package \textbf{mvordnorm}
implements estimation of this model using composite likelihood methods
and is available for download from Github. We present two use-cases in
the area of risk management to illustrate our approach. The first
illustration uses a data set containing defaults, ratings and CDS
spreads for US listed companies. Moreover, we have also identified this
framework to be useful for the modelling of ESG ratings and continuous
ESG scores, which have been show to have low comparability and
integration (Berg, Koelbel, and Rigobon 2019), so the second application
relies on a data set containing ESG ratings from three data providers:
RepRisk, Sustainalytics and Refinitiv. As covariates we use financial
variables identified by the literature to be relevant in explaining the
above mentioned measures. In our exercises we show promising results
related to the performance of the joint model for both the credit risk
and the ESG risk application.

The paper is organized as follows: Section 2 presents the joint model
and Section 3 discusses the estimation procedure using composite
likelihood methods. The software implementation is described in Section
4. The modeling approach is illustrated on two use-cases in Section 5.
Section 6 concludes.

\section{Model}\label{sect:model}

We propose a joint model of continuous and ordinal variables. Assume we
have \(q_n\) continuous responses and \(q_o\) ordinal responses (in
total \(q=q_n+q_o\)).

For the continuous responses
\(\boldsymbol y^n_{1},\ldots, \boldsymbol y^n_{q_n}\) we assume the
following regression model: \[
y^n_{ij} = \beta_{0j} + \bm\beta_j^\top \bm x_i + \sigma_j \epsilon^n_{ij}
\] for all observations \(i=1,\ldots,n\) and \(j=1,\ldots,q_n\) where
\(\bm x_i\) is a \((p\times 1)\) vector of covariates, \(\bm\beta_j\) is
a \((p\times 1)\) vector of response-specific regression coefficients,
\(\sigma_j\) is a response-specific scale parameter and
\(\epsilon^n_{ij}\) is a mean-zero error term.

For the ordinal variables
\(\boldsymbol y^o_{1},\ldots, \boldsymbol y^o_{q_o}\) a regression model
is assumed on the latent variable \(\tilde{\bm y}\) underlying the
ordinal observations: \[
\tilde y^o_{ij} =\bm \beta_j^\top \bm x_i + \epsilon^o_{ij},
\] where the observed responses are obtained by slotting the continuous
latent variables using appropriate threshold parameters. \[
y^o_{ij} = r \Rightarrow \theta_{j, r-1}\leq \tilde y^o_{ij}\leq \theta_{j, r}, \quad r\in \{1, \ldots, K_j\} \quad  -\infty\equiv\theta_{j, 0} <\theta_{j, 1}<\ldots<\theta_{j, K}\equiv \infty,
\] where \(K_j\) denotes the number of categories for ordinal response
\(j=1, \ldots, q_o\).

The dependence is captured by assuming that the errors of the continuous
and the latent variables of the ordinals come from a multivariate normal
distribution: \[
(\boldsymbol\epsilon^o_{i}, \boldsymbol\epsilon^n_{ij})^\top \sim N(\bm 0, R)
\] with mean zero and correlation matrix \(R\). Note that the matrix is
restricted to have diagonal elements equal to one due to identifiability
restrictions for the ordinal variables.

\section{Pairwise likelihood}\label{sect:pl}

Assume we collect all \(q\) responses in a matrix \(Y\) of dimension
\((n\times q)\) and that there are no missing values in the response
matrix. The pairwise log-likelihood is given by the sum of the bivariate
likelihoods over all pairs of responses \(k=1,\ldots,q\) and \(l > k\):
\[
p\ell(\Theta; Y, X) =\sum_{i=1}^n \sum_{k<l} \ell(\Theta; y_{ik}, y_{il}).
\] The bivariate log-likelihoods can be split into three cases,
depending on the type of the responses \(k\) and \(l\):

\begin{itemize}
\item
  Case 1: both responses are ordinal
\item
  Case 2: both responses are normal
\item
  Case 3: one response is ordinal, one normal
\end{itemize}

For \emph{Case 1}, the bivariate log-likelihood is given by the
bivariate probability \(\Pr(y_{ik}=r_{ik}, y_{il} = r_{il})\): \[
\ell(\Theta; y_{ik}, y_{il})=\Pr(y_{ik}=r_{ik}, y_{il} = r_{il}) = 
\int_{\theta_{k, r_{ik}-1}}^{\theta_{k, r_{ik}}} 
\phi_2(\tilde y_{ik} -\bm \beta_k^\top \bm x_i, \tilde y_{il} - \bm\beta_k^\top \bm x_i; \rho_{kl}) d\tilde y_{ik}d\tilde y_{il},
\] where \(\rho_{kl}\) is the correlation coefficient between responses
\(k\) and \(l\) and \(\phi_2\) is the bivariate standard normal
probability density function. By denoting
\(U_{i,k} = \theta_{k, r_{ik}} -\bm \beta_k^\top \bm x_i\) ,
\(L_{i,k} = \theta_{k, r_{ik}-1} - \bm \beta_k^\top \bm x_i\),
\(U_{i,l} = \theta_{k, r_{il}} - \bm \beta_l^\top \bm x_i\),
\(L_{i,l} = \theta_{k, r_{ik}-1} - \bm \beta_l^\top \bm x_i\), the
bivariate probability is equal to: \[
\Pr(y_{ik}=r_{ik}, y_{il} = r_{il})=\Phi_2\left(U_{i,k}, U_{i,l};  \rho_{kl} \right)-\Phi_2\left(L_{i,k}, U_{i,l};  \rho_{kl} \right)-\Phi_2\left(U_{i,k}, L_{i,l};  \rho_{kl} \right)
+\Phi_2\left(L_{i,k}, L_{i,l};  \rho_{kl} \right)
\] where \(\Phi_2\) is the bivariate standard normal cumulative
distribution function.

For \emph{Case 2}, where both responses are continuous, the bivariate
log-likelihood is given by the bivariate standard normal density: \[
\ell(\Theta; y_{ik}, y_{il}) = f(y_{ik}, y_{il}|\Theta)= \phi_2\left(\frac{y_{ik} - \beta_{0k} -\bm\beta_k^\top \bm x_i}{\sigma_k}, \frac{y_{il} - \beta_{0l} -\bm\beta_l^\top \bm x_i}{\sigma_l}; \rho_{kl}) \right)
\]

Finally, for \emph{Case 3}, where one response is ordinal and one
response is continuous, the bivariate log-likelihood is given by the
product of a conditional probability and the density of the continuous
response. Assuming that the \(k\)-th response is ordinal and the
\(l\)-th is continuous, we have: \[
\ell(\Theta; y_{ik}, y_{il}) = \Pr(y_{ik}=r_{ik} | y^n_{il}) f(y_{il})
\] where \(f(y_{il})\) is the marginal standard normal pdf
\(\phi\left(\frac{y^n_{il} - \beta_{0l} -\bm\beta_l^\top \bm x_i}{\sigma_l} \right)\).
The conditional probability is given by : \[
\Pr(y_{ik}=r_{ik} | y_{il}) = \Pr(\theta_{k, r_{ik}-1}\leq \tilde y_{ik}\leq \theta_{k, r_{ik}} | y_{il}).
\] Using the fact that \(\tilde y_{ik}\) and \(y_{il}\) are bivariate
normals, we can easily derive the conditional distribution \[
\tilde y_{ik} | y_{il} \sim N\left(\underbrace{\bm\beta_k^\top \bm x_i + \frac{\rho_{kl}}{\sigma_l}(y_{il} - \beta_{0l} -\bm\beta_l^\top \bm x_i)}_{\mu_c}, \underbrace{(1 - \rho_{kl}^2)}_{\sigma_c^2}\right)
\] and the corresponding conditional probability: \[
\Pr(y_{ik}=r_{ik} | y_{il})=
\int_{\theta_{k, r_{ik}-1}}^{\theta_{k, r_{ik}}} \phi\left(\frac{\tilde y_{ik} - \mu_c}{\sigma_c}\right) d\tilde y_{ik} = 
\Phi\left(\frac{\theta_{k, r_{ik}} - \mu_c}{\sigma_c}\right)-\Phi\left(\frac{\theta_{k, r_{ik}-1} - \mu_c}{\sigma_c}\right).
\]

The standard errors of the parameters we computed using the Godambe
information matrix and model comparison can be performed using modified
version of Akaike and Bayesian information criteria (for more details
see Hirk, Hornik, and Vana 2020)

Missing values

Note that the framework is easily extended to allow for the presence of
missing values in the responses. For each observation \(i\), in the
calculation of the pairwise log-likelihood, we consider all pairs of
\emph{observed responses}. In case for some observations only one
response is available, we consider the likelihood of the univariate
observed response. For the continuous responses this is the standard
normal pdf
\(\phi\left(\frac{y^n_{ik} - \beta_{0k} -\bm\beta_k^\top \bm x_i}{\sigma_l} \right)\)
while for the ordinal responses we have the univariate probability
\(\Pr(y_{ik}=r_{ik})=\Phi(U_{i,k}) - \Phi(L_{i,k})\)

\section{Software implementation}\label{sect:soft}

The model is implemented in package \textbf{mvordnorm}. The main
function for fitting the models is \texttt{mvordnorm()}.

\begin{Shaded}
\begin{Highlighting}[]
\FunctionTok{mvordnorm}\NormalTok{(formula, data, }\AttributeTok{response\_types =} \ConstantTok{NULL}\NormalTok{, na.action,}
  \AttributeTok{contrasts =} \ConstantTok{NULL}\NormalTok{, }\AttributeTok{control =} \FunctionTok{mvordnorm.control}\NormalTok{(), ...)}
\end{Highlighting}
\end{Shaded}

The main arguments are:

\begin{itemize}
\item
  \texttt{formula}: a formula object of class \texttt{"Formula"}, as
  implemented in the \textbf{Formula} package (Zeileis and Croissant
  2010), which makes it easy to specify multiple responses, especially
  if these are not numeric or of mixed type.
\item
  \texttt{data}: a data frame which contains the different responses in
  separate columns.
\item
  \texttt{response\_types}: a (named) vector of characters with length
  equal to the number of responses. Each element of the vector is either
  \texttt{"gaussian"} or \texttt{"ordinal"}.
\item
  \texttt{na.action}: a function which indicates what should happen when
  the data contain NAs.
\item
  \texttt{contrasts}: an optional list. See the \texttt{contrasts.arg}
  of \texttt{model.matrix.default}.
\item
  \texttt{control}: list of parameters for controlling the fitting
  process such as the general purpose solver to be used or whether
  standard errors should be calculated.
\end{itemize}

For illustration purposes in section we use a worked example based on a
simulated data set consisting of 1000 subjects for which two multiple
ordinal responses (\texttt{y1} and \texttt{y2}), two continuous
responses (\texttt{z1} and \texttt{z2}) and three covariates
(\texttt{x1}, \texttt{x2} and \texttt{x3}) are available. The ordinal
responses each have three categories labeled with 1, 2 and 3.

\begin{Shaded}
\begin{Highlighting}[]
\FunctionTok{library}\NormalTok{(}\StringTok{"mvordnorm"}\NormalTok{)}
\FunctionTok{data}\NormalTok{(}\StringTok{"data\_toy"}\NormalTok{, }\AttributeTok{package =} \StringTok{"mvordnorm"}\NormalTok{)}
\FunctionTok{dim}\NormalTok{(data\_toy)}
\end{Highlighting}
\end{Shaded}

\begin{verbatim}
## [1] 1000    7
\end{verbatim}

\begin{Shaded}
\begin{Highlighting}[]
\FunctionTok{head}\NormalTok{(data\_toy)}
\end{Highlighting}
\end{Shaded}

\begin{verbatim}
##   y1 y2         z1           z2         X1          X2          X3
## 1  3  2  0.9696782  2.805655841  0.5855288  1.67751179 -0.60784111
## 2  2  2 -1.4973329  0.306208073  0.7094660  0.07947405  1.07622314
## 3  2  2 -0.6577188 -0.002179144 -0.1093033 -0.85642750 -0.57642579
## 4  1  1 -4.3087039 -2.588533195 -0.4534972 -0.77877729  1.09862636
## 5  1  2 -2.4160105 -1.110698320  0.6058875 -0.38093608  1.40734169
## 6  1  1 -5.0378041 -4.184419290 -1.8179560 -1.89735834  0.03665615
\end{verbatim}

In order to estimate the model using the CG solver in \textbf{optimx}
(Nash 2014), we have the following function call:

\begin{Shaded}
\begin{Highlighting}[]
\NormalTok{fit }\OtherTok{\textless{}{-}} \FunctionTok{mvordnorm}\NormalTok{(}\StringTok{"y1 + y2 + z1 + z2 \textasciitilde{} 0 + X1 + X2 + X3"}\NormalTok{, }\AttributeTok{data =}\NormalTok{ data\_toy,}
          \AttributeTok{response\_types =} \FunctionTok{c}\NormalTok{(}\StringTok{"ordinal"}\NormalTok{, }\StringTok{"ordinal"}\NormalTok{,}\StringTok{"gaussian"}\NormalTok{, }\StringTok{"gaussian"}\NormalTok{),}
          \AttributeTok{control =} \FunctionTok{mvordnorm.control}\NormalTok{(}\AttributeTok{se =} \ConstantTok{TRUE}\NormalTok{, }\AttributeTok{solver =} \StringTok{"CG"}\NormalTok{))}
\end{Highlighting}
\end{Shaded}

Note that the formula specifies no intercept is to be estimated, this is
however due to the intercept not being identifiable in ordinal
regression. For the normal variables, an intercept will be computed by
default. This peculiarity of the implementation should be improved in
future versions of the package, to allow the user to specify outcome
specific intercepts.

The \texttt{summary} method produces an output which is similar to the
one of most regression models, and contains information on the
thresholds for ordinal responses and intercepts for the continuous
response, on the outcome specific regression coefficients, on the scale
parameters for the continuous variables and the correlation matrix
\(R\).

\begin{Shaded}
\begin{Highlighting}[]
\FunctionTok{summary}\NormalTok{(fit)}
\end{Highlighting}
\end{Shaded}

\begin{verbatim}
## Formula: y1 + y2 + z1 + z2 ~ 0 + X1 + X2 + X3 | 1
## <environment: 0x11f38e8c8>
## 
## nunits ndim    logLik
##   1000    4 -11565.04
## 
## Thresholds:
##         Estimate Std. Error z value  Pr(>|z|)    
## y1 1|2 -1.006292   0.074042 -13.591 < 2.2e-16 ***
## y1 2|3  1.025590   0.070709  14.504 < 2.2e-16 ***
## y2 1|2 -1.940256   0.090877 -21.351 < 2.2e-16 ***
## y2 2|3  2.000534   0.090580  22.086 < 2.2e-16 ***
## 
## Intercept for normals 
##           Estimate Std. Error z value  Pr(>|z|)    
## beta0.z1 -1.000729   0.032022 -31.251 < 2.2e-16 ***
## beta0.z2  0.939933   0.063967  14.694 < 2.2e-16 ***
## 
## Coefficients:
##       Estimate Std. Error  z value Pr(>|z|)    
## y1X1  1.980493   0.095712  20.6922   <2e-16 ***
## y2X1  1.978757   0.083675  23.6482   <2e-16 ***
## z1X1  2.022099   0.033480  60.3977   <2e-16 ***
## z2X1  2.069122   0.066792  30.9786   <2e-16 ***
## y1X2 -0.017407   0.045554  -0.3821   0.7024    
## y2X2 -0.035601   0.045292  -0.7860   0.4319    
## z1X2 -0.018743   0.031548  -0.5941   0.5524    
## z2X2 -0.070176   0.063849  -1.0991   0.2717    
## y1X3 -1.926444   0.094785 -20.3243   <2e-16 ***
## y2X3 -2.005459   0.087264 -22.9815   <2e-16 ***
## z1X3 -2.010065   0.035192 -57.1169   <2e-16 ***
## z2X3 -1.977850   0.069517 -28.4513   <2e-16 ***
## 
## Standard deviation of the Gaussian response variables:
##          Estimate Std. Error z value  Pr(>|z|)    
## sigma.z1 0.992114   0.021779  45.554 < 2.2e-16 ***
## sigma.z2 1.985102   0.044032  45.083 < 2.2e-16 ***
## 
## Correlation params:
##             Estimate Std. Error z value  Pr(>|z|)    
## corr_y1_y2 0.6363705  0.0502925  12.653 < 2.2e-16 ***
## corr_y1_z1 0.7799013  0.0197699  39.449 < 2.2e-16 ***
## corr_y1_z2 0.6542121  0.0299041  21.877 < 2.2e-16 ***
## corr_y2_z1 0.9188427  0.0130324  70.504 < 2.2e-16 ***
## corr_y2_z2 0.8011830  0.0222478  36.012 < 2.2e-16 ***
## corr_z1_z2 0.8962955  0.0061244 146.349 < 2.2e-16 ***
\end{verbatim}

The \texttt{summary} method produces an output which is similar to the
one of most regression models.

If a model should be fit to data containing missing values, the
\texttt{na.action} argument should be set to \texttt{na.pass}. We
introduce some NAs in \texttt{data\_toy}:

\begin{Shaded}
\begin{Highlighting}[]
\NormalTok{data\_toy}\SpecialCharTok{$}\NormalTok{y1[}\FunctionTok{sample}\NormalTok{(}\DecValTok{1}\SpecialCharTok{:}\FunctionTok{nrow}\NormalTok{(data\_toy), }\DecValTok{20}\NormalTok{)] }\OtherTok{\textless{}{-}} \ConstantTok{NA}
\NormalTok{data\_toy}\SpecialCharTok{$}\NormalTok{y2[}\FunctionTok{sample}\NormalTok{(}\DecValTok{1}\SpecialCharTok{:}\FunctionTok{nrow}\NormalTok{(data\_toy), }\DecValTok{20}\NormalTok{)] }\OtherTok{\textless{}{-}} \ConstantTok{NA}
\end{Highlighting}
\end{Shaded}

The function call is:

\begin{Shaded}
\begin{Highlighting}[]
\NormalTok{fit\_with\_NAs }\OtherTok{\textless{}{-}} \FunctionTok{mvordnorm}\NormalTok{(}\StringTok{"y1 + y2 + z1 + z2 \textasciitilde{} 0 + X1 + X2 + X3"}\NormalTok{, }\AttributeTok{data =}\NormalTok{ data\_toy,}
                          \AttributeTok{na.action =}\NormalTok{ na.pass,}
                          \AttributeTok{response\_types =} \FunctionTok{c}\NormalTok{(}\StringTok{"ordinal"}\NormalTok{, }\StringTok{"ordinal"}\NormalTok{,}\StringTok{"gaussian"}\NormalTok{, }\StringTok{"gaussian"}\NormalTok{),}
                          \AttributeTok{control =} \FunctionTok{mvordnorm.control}\NormalTok{(}\AttributeTok{se =} \ConstantTok{TRUE}\NormalTok{, }\AttributeTok{solver =} \StringTok{"CG"}\NormalTok{))}
\FunctionTok{summary}\NormalTok{(fit\_with\_NAs)}
\end{Highlighting}
\end{Shaded}

\begin{verbatim}
## Formula: y1 + y2 + z1 + z2 ~ 0 + X1 + X2 + X3 | 1
## <environment: 0x138ad5140>
## 
## nunits ndim   logLik
##   1000    4 -11375.8
## 
## Thresholds:
##         Estimate Std. Error z value  Pr(>|z|)    
## y1 1|2 -1.001616   0.074587 -13.429 < 2.2e-16 ***
## y1 2|3  1.035621   0.071466  14.491 < 2.2e-16 ***
## y2 1|2 -1.951031   0.091063 -21.425 < 2.2e-16 ***
## y2 2|3  2.007763   0.090616  22.157 < 2.2e-16 ***
## 
## Intercept for normals 
##           Estimate Std. Error z value  Pr(>|z|)    
## beta0.z1 -0.999231   0.032208 -31.024 < 2.2e-16 ***
## beta0.z2  0.941897   0.064413  14.623 < 2.2e-16 ***
## 
## Coefficients:
##        Estimate Std. Error  z value Pr(>|z|)    
## y1X1  1.9892361  0.0968380  20.5419   <2e-16 ***
## y2X1  1.9852961  0.0833405  23.8215   <2e-16 ***
## z1X1  2.0188906  0.0336378  60.0186   <2e-16 ***
## z2X1  2.0627172  0.0672441  30.6751   <2e-16 ***
## y1X2 -0.0073997  0.0458717  -0.1613   0.8718    
## y2X2 -0.0174270  0.0467789  -0.3725   0.7095    
## z1X2 -0.0159245  0.0317781  -0.5011   0.6163    
## z2X2 -0.0635339  0.0643731  -0.9870   0.3237    
## y1X3 -1.9253368  0.0951997 -20.2242   <2e-16 ***
## y2X3 -2.0272346  0.0876712 -23.1231   <2e-16 ***
## z1X3 -2.0072877  0.0353274 -56.8196   <2e-16 ***
## z2X3 -1.9721185  0.0699664 -28.1867   <2e-16 ***
## 
## Standard deviation of the Gaussian response variables:
##          Estimate Std. Error z value  Pr(>|z|)    
## sigma.z1 0.993199   0.021895  45.362 < 2.2e-16 ***
## sigma.z2 1.989667   0.044376  44.837 < 2.2e-16 ***
## 
## Correlation params:
##             Estimate Std. Error z value  Pr(>|z|)    
## corr_y1_y2 0.6517810  0.0500851  13.014 < 2.2e-16 ***
## corr_y1_z1 0.7830532  0.0196507  39.849 < 2.2e-16 ***
## corr_y1_z2 0.6499493  0.0305379  21.283 < 2.2e-16 ***
## corr_y2_z1 0.9189966  0.0134847  68.151 < 2.2e-16 ***
## corr_y2_z2 0.8035944  0.0221505  36.279 < 2.2e-16 ***
## corr_z1_z2 0.8964118  0.0061399 145.997 < 2.2e-16 ***
\end{verbatim}

\section{Empirical analysis}\label{sect:empirical}

In the empirical analysis we present two use-cases. The first use-case
uses data containing default, credit ratings and CDS spreads for a
sample of US companies from 2003-2013. For the second use-case we
collect data from three ESG providers: Refinitiv, Sustainalytics and
RepRisk and build a joint model of their ESG scores and ratings.

\subsection{Credit ratings, failure and CDS
data}\label{credit-ratings-failure-and-cds-data}

We construct a sample of Compustat/CRSP companies from US over the
period 2003--2013. We use S\&P long-term issuer credit ratings from the
Compustat-Capital IQ Credit Ratings database as well as issuer credit
ratings from Moody's The failure indicator is constructed based on the
default data from the UCLA-LoPucki Bankruptcy Research Database and the
Mergent issuer default file. A binary failure indicator is constructed
in the following way: a default is recorded in a year if a firm filed
for bankruptcy under Chapter 7 or Chapter 11 or the firm receives a
default rating from one of the CRAs in the year following the rating
observation. Finally, we obtain CDS pricing data from IHS Markit via
Wharton research data services (WRDS). We only keep CDS with a primary
coupon, with a short duration of 1 year and merge the end-of-year
spreads with the end-of-year ratings. Due to the lack of common unique
identifiers to be used for merging and due to coverage issues we only
observe CDS spreads over the period 2008-2013.

The covariates are built using the Compustat and CRSP databases together
with the corresponding linking files available via WRDS. We exclude
financial, utility and real estate firms from the data set. The
end-of-year ratings and 1-year CDS spread are merged to the financial
ratios on a calendar year basis.

In the analysis we employ the variables proposed in (Alp 2013) which
include interest coverage, EBITDA to sales, long-term debt to total
assets, liabilities to assets, firm size (as percentile of firm's market
capitalization\\
in the distribution of the NYSE stocks capitalization) and whether it is
a dividend payer, idiosyncratic (SIGMA) and systematic risk (BETA),
retained earnings to assets, R\&D expenses to assets, cash- and tangible
assets-to-assets, market to book assets ratio. All ratios are winsorized
at 99\% and, if negative values are present, at 1\%.

\begin{figure}
\includegraphics[width=1\linewidth]{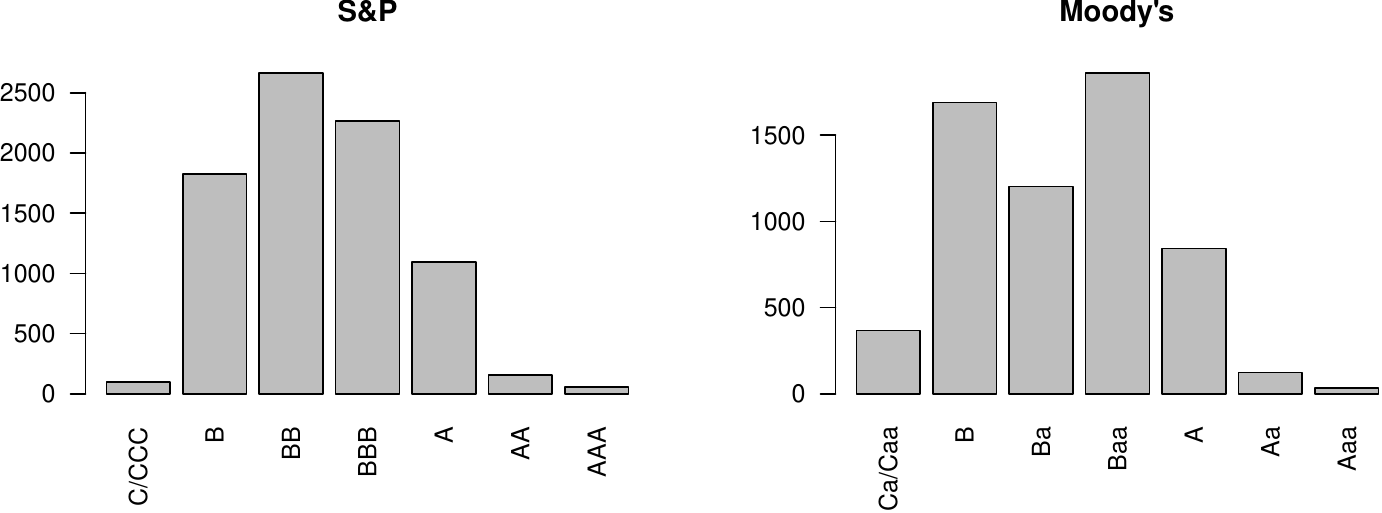} \caption{\label{fig:rat} Rating distribution for the three rating agencies over the whole sample 2003-2013}\label{fig:unnamed-chunk-9}
\end{figure}

After eliminating missing values in the covariates, The merged sample
contains 1477 firms and firm-year observations. The rating distribution
(for the aggregated ratings without modifiers) is shown Figure
\ref{fig:rat}. We observed a mode in the BB and Baa classes rating
agencies, with few observations falling into the best rating classes.

\begin{figure}

{\centering \includegraphics[width=0.6\linewidth]{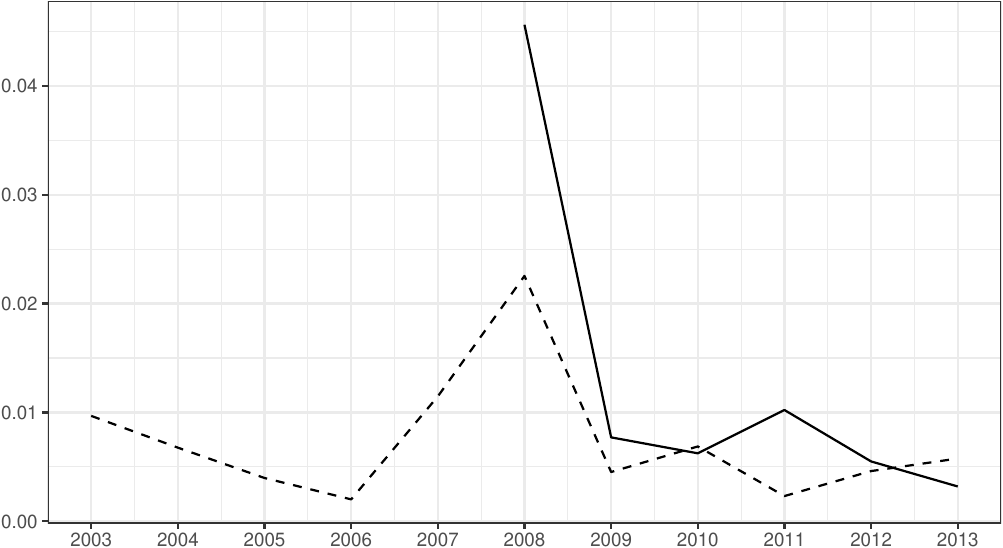} 

}

\caption{\label{fig:cdsyears} Average CDS spread (full) and average default rate (dashed) over sample period 2003-2013. Note that the CDS data starts in 2008.}\label{fig:unnamed-chunk-10}
\end{figure}

The average CDS spreads for each year are presented in Figure
\ref{fig:cdsyears}.

We estimate a joint model for the four responses: S\&P and Moody's
ratings, failure indicator and the log of the CDS spread. We standardize
all covariates to have mean zero and variance one. Therefore, the
regression coefficients for one outcome can be interpreted in terms of
importance.

\begin{figure}

{\centering \includegraphics[width=0.5\linewidth]{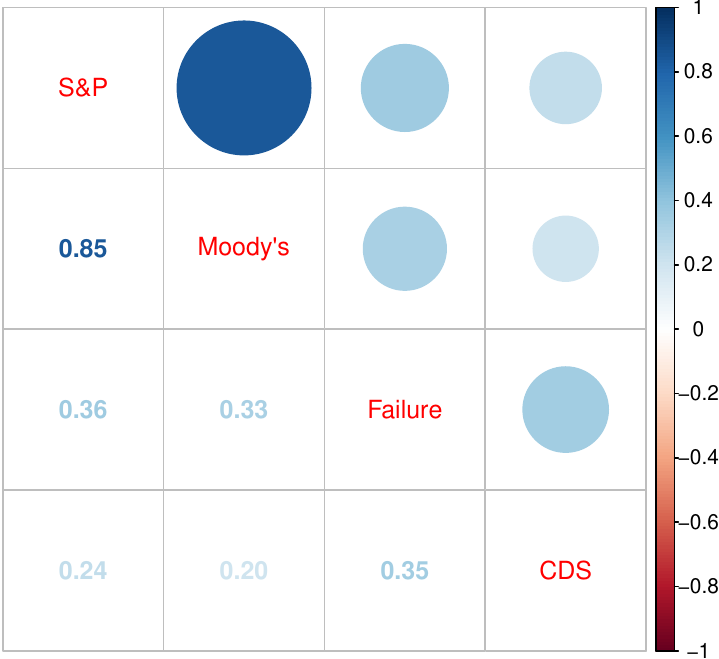} 

}

\caption{\label{fig:corrcds}Estimated correlation matrix among the response variables in the credit risk application.}\label{fig:unnamed-chunk-12}
\end{figure}

Table \ref{tab:coef} presents the estimated regression coefficients for
each response while Figure \ref{fig:corrcds} illustrates the estimated
correlation matrix among the responses. We observe that almost all the
variables have the expected sign. The coefficients for the rating
dimensions are rather similar, with size being the most important
variable.

\begin{table}[ht]
\centering
\begin{tabular}{lrrrr}
  \toprule
 & \multicolumn{1}{c}{S\&P} & \multicolumn{1}{c}{Moody's} & \multicolumn{1}{c}{Failure} & \multicolumn{1}{c}{CDS} \\ 
  \midrule
Int.coverage & $-$0.1009 (0.0155)*** & $-$0.1130 (0.0249)*** & $-$0.1481 (0.0823).\phantom{**} & $-$0.0314 (0.0546)\phantom{***} \\ 
  EBITDA/sale & $-$0.0434 (0.0229).\phantom{**} &  0.0120 (0.0503)\phantom{***} & $-$0.0411 (0.0403)\phantom{***} &  0.3042 (0.1442)*\phantom{**} \\ 
  LT debt/assets &  0.3546 (0.0464)*** &  0.5264 (0.0493)*** & $-$0.0513 (0.0588)\phantom{***} & $-$0.3320 (0.1157)*** \\ 
  Liab/assets & $-$0.0235 (0.0468)\phantom{***} & $-$0.1842 (0.0482)*** &  0.2503 (0.0661)*** &  0.4466 (0.1101)*** \\ 
  SIZE & $-$1.1944 (0.0216)*** & $-$1.2541 (0.0231)*** & $-$0.4543 (0.0450)*** & $-$0.4198 (0.0382)*** \\ 
  SIGMA &  0.5425 (0.0175)*** &  0.4624 (0.0202)*** &  0.6965 (0.0265)*** &  1.2924 (0.0393)*** \\ 
  BETA &  0.2245 (0.0138)*** &  0.2047 (0.0157)*** &  0.0133 (0.0220)\phantom{***} & $-$0.1235 (0.0350)*** \\ 
  Market/book & $-$0.0524 (0.0137)*** & $-$0.0946 (0.0145)*** & $-$0.0949 (0.0326)*** & $-$0.1212 (0.0233)*** \\ 
  R\&D/assets & $-$0.1231 (0.0191)*** & $-$0.0772 (0.0232)*** &  0.0677 (0.0473)\phantom{***} & $-$0.0290 (0.0414)\phantom{***} \\ 
  Ret.earn./assets & $-$0.7670 (0.0157)*** & $-$0.9647 (0.0295)*** & $-$0.0107 (0.0297)\phantom{***} & $-$0.1170 (0.0766)\phantom{***} \\ 
  Capex/assets &  0.1101 (0.0183)*** &  0.0997 (0.0203)*** &  0.0490 (0.0300)\phantom{***} &  0.1326 (0.0547)*\phantom{**} \\ 
  Cash/assets &  0.1836 (0.0190)*** &  0.1765 (0.0209)*** & $-$0.4176 (0.0503)*** &  0.0365 (0.0355)\phantom{***} \\ 
  Tang./assets &  0.0423 (0.0188)*\phantom{**} &  0.0875 (0.0208)*** &  0.1260 (0.0313)*** & $-$0.0671 (0.0420)\phantom{***} \\ 
   \bottomrule
\end{tabular}
\caption{This table displays the estimated regression coefficients for the four outcomes: S\&P ratings, Moody's ratings, failure indicator and the log CDS spread.} 
\label{tab:coef}
\end{table}

\subsection{ESG ratings}\label{esg-ratings}

ESG ratings and scores are observed to be rather inhomogeneous among the
different ratings sources. There are currently more than 140 providers
-- well known providers include Refinitiv, Sustainalytics, RepRisk, MSCI
ESG Research, S\&P Global ESG Score, CDP Climate, Water and Forest
Scores, Bloomberg ESG Disclosures Scores, ISS Ratings and Rankings and
ESGI.

In this paper we collect historical ESG ratings and scores from
Refinitiv, Sustainalytics and RepRisk for the constituents of the S\&P
500 index in the year 2017, with access provided by our host
institution. We obtain financial variables from Refinitiv which we use
as covariates in the analysis. We merge ESG indicators from all
providers based on tickers. From RepRisk we obtain ratings on a 10-point
ordinal scale. From Refinitiv and Sustainalytics we obtains ESG scores
on a continuous scale.

We employ 17 covariates related to financial performance, including
gross profit margin, EBITDA margin, asset and fixed asset turnover,
income tax rate, liquidity ratios, return on equity and return on
capital. After eliminating missing values in the covariates the data set
contains 2451.

\begin{figure}
\includegraphics[width=1\linewidth]{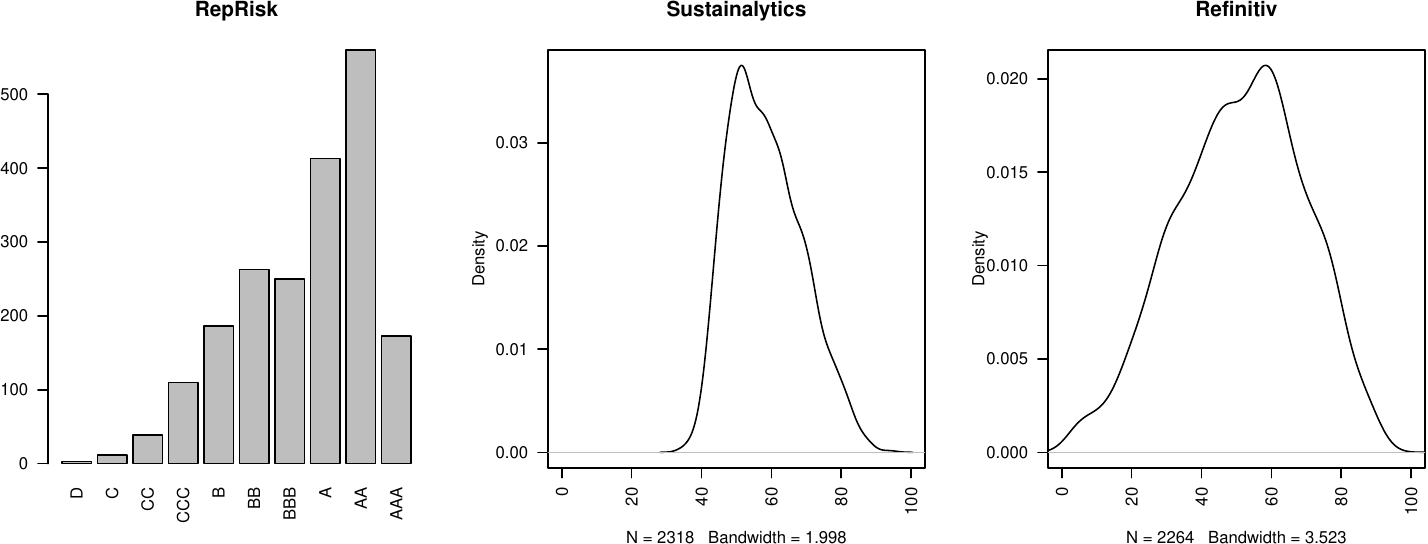} \caption{\label{fig:esgrat} Rating distribution for the three rating agencies over the whole sample 2003-2013}\label{fig:unnamed-chunk-14}
\end{figure}

Figure \ref{fig:esgrat} presents the distribution of the ESG scores and
ratings for the three providers.

\begin{figure}

{\centering \includegraphics[width=0.5\linewidth]{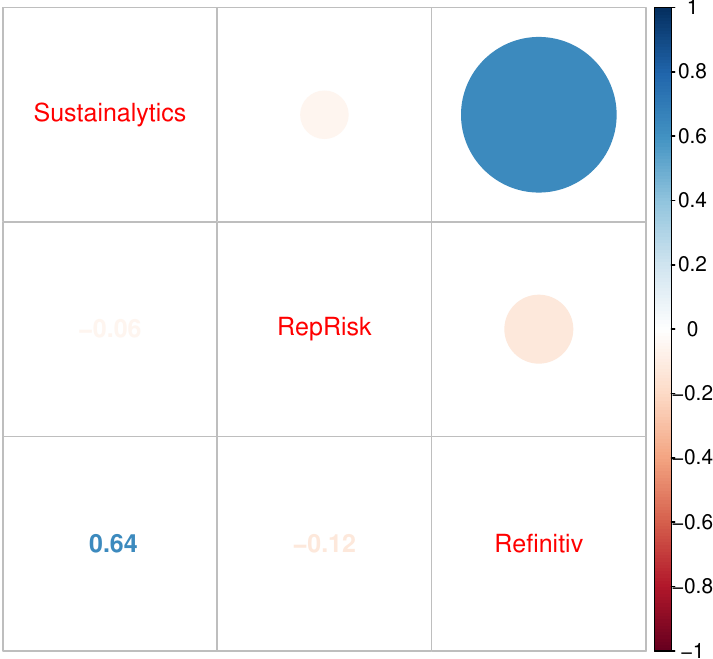} 

}

\caption{\label{fig:corresg}Estimated correlation matrix among the response variables in the ESG application.}\label{fig:unnamed-chunk-15}
\end{figure}

We estimate the joint model of three responses (one ordinal and two
continuous). Figure \ref{fig:corresg} shows the estimated correlation
among the three responses. The correlations among the responses are
generally low, indicating a low degree of agreement among the providers
(in line with Berg, Koelbel, and Rigobon 2019). It is interesting to
note that the there is a negative correlation among the RepRisk ratings
and the score of the other two providers. It is to be noted that from
the documentation of the providers, it is difficult to grasp whether the
score provided by Sustainalytics indicates the riskiness or how well the
company is doing in terms of ESG metrics. Also, the providers changed
the scoring and rating methodology in the previous years, which
constitutes an additional issue when interpreting the results.

The regression coefficients of the employed covariates (after performing
a stepwise elimination of the non-significant variable, are presented in
Table \ref{tab:coefesg}. We observe that the financial covariates have
low explanatory power for the ESG scores, with current ratio being the
only significant variable for Refinitiv's combined ESG score. In
addition to the current ratio, the rate of earning retention is also
relevant in explaining the total ESG score of Sustainalytics. It seems
that for this sample, the covariates based on financial statements have
low explanatory power for these providers. Further, more extensive
experiments should however be performed.

\begin{table}[ht]
\centering
\begin{tabular}{lrrr}
  \toprule
 & \multicolumn{1}{c}{Sustainalytics} & \multicolumn{1}{c}{RepRisk} & \multicolumn{1}{c}{Refinitiv} \\ 
  \midrule
Asset turnover & $-$0.0732 (0.2783)\phantom{***} &  0.0631 (0.0229)*** & $-$0.2564 (0.4833)\phantom{***} \\ 
  Assets/Comm.equity & $-$0.0746 (0.3458)\phantom{***} & $-$0.0424 (0.0156)*** &  0.0395 (0.7331)\phantom{***} \\ 
  Earnings Retention Rate & $-$0.4608 (0.1417)*** & $-$0.1155 (0.0575)*\phantom{**} & $-$1.1617 (0.7917)\phantom{***} \\ 
  Current.Ratio & $-$1.0435 (0.2882)*** &  0.0646 (0.0241)*** & $-$1.8822 (0.4698)*** \\ 
  Fixed asset turnover &  0.2984 (1.0895)\phantom{***} &  0.9926 (0.1076)*** & $-$0.5384 (2.4630)\phantom{***} \\ 
   \bottomrule
\end{tabular}
\caption{This table displays the estimated regression coefficients for the three outcomes in the ESG application: Sustainalytics total ESG scores, RepRisk ratings and 
             Refinitiv ESG combined scores.} 
\label{tab:coefesg}
\end{table}

\section{Conclusion}\label{sect:concl}

We propose a framework for jointly modelling continuous and ordinal
responses by assuming that the latent variables underlying the ordinal
observations and the observed continuous responses follow a multivariate
normal distribution. The estimation is performed by maximizing the
pairwise likelihood. We exemplify the framework on two use-cases: a
credit risk application which jointly looks at credit ratings from
Moody's and S\&P, failures and CDS spreads for listed companies in the
US over 2003-2013, and an ESG application where ESG ratings and scores
for large US companies in the year 2017.

Methodological extensions include consideration of other error
distributions such as multivariate Student-\(t\). Moreover, for the
empirical analysis it would be of interest to further investigate how
one can incorporate both ESG and credit risk measures in such a joint
model and what insights can be derived from such an approach. Finally,
implementing an automatic model selection procedure for the multivariate
model is also on the research agenda of the authors.

\paragraph{Acknowledgments}\label{acknowledgments}

This research was supported by funds of the Oesterreichischen
Nationalbank (Austrian Central Bank, Anniversary Fund, project number:
18482 ``Multivariate ordinal regression models for enhanced credit risk
modeling'\,').

\section{References}\label{references}

\phantomsection\label{refs}

\newpage

\section{Appendix}\label{appendix}

\subsection{Derivation of analytical standard
errors}\label{derivation-of-analytical-standard-errors}

\subsubsection{Univariate case}\label{univariate-case}

If for some observations only one response \(j\) is observed, then we
have two cases: either an ordinal response, or a continuous response.

\paragraph{Ordinal response}\label{ordinal-response}

The negative log-likelihood is: \[
\textrm{neglog}\mathcal L^j_{i} = -\log\left(\Phi\left( \theta_{j, r_{ij}} - \boldsymbol \beta_j^\top \boldsymbol x_i\right)-\Phi\left( \theta_{j, r_{ij}-1} - \boldsymbol \beta_j^\top \boldsymbol x_i\right)\right)
\] Assume
\(\tilde{\bm x}^\text{upper}_{i,j} = (0,1,..,-\boldsymbol x_i)^\top\)
and \(\tilde{\bm x}^\text{lower}_{i,j} = (1,0,..,-\bm x_i)^\top\) (i.e.,
we make a dummy matrix of dimension \(K_j\) where for the lower
predictors we eliminate the first column and for the upper predictors we
eliminate the last column) and
\(\boldsymbol \psi_j=(\boldsymbol \theta_j, \boldsymbol \beta_j)^\top\).
The relevant derivatives are:

\[
\frac{\partial\textrm{neglog}\mathcal L^j_{i}}{\boldsymbol \psi_j}=-\frac{1}{\Phi\left(\boldsymbol\psi_j^\top
\tilde{\bm x}^\text{upper}_{i,j}\right)-\Phi\left(\boldsymbol\psi_j^\top 
\tilde{\bm x}^\text{lower}_{i,j}\right)}(\phi(\boldsymbol\psi_j^\top \tilde{\bm x}^\text{upper}_i)\tilde{\bm x}^\text{upper}_{i,j} - \phi(\boldsymbol\psi_j^\top \tilde{\bm x}^\text{lower}_{i,j})\tilde{\bm x}^\text{lower}_{i,j})
\]

\paragraph{Gaussian responses}\label{gaussian-responses}

The negative log-likelihood is: \[
\textrm{neglog}\mathcal L^j_{i} = -\log\left(\frac{1}{\sqrt{2\pi\sigma^2_j}}\exp\left(-\frac{(y_i - \boldsymbol\beta_{j}^{*\top} \boldsymbol x^*_i)^2}{2\sigma_j^2}\right)\right) = 0.5\log(2\pi)+\log(\sigma_j) + \frac{(y_i - \beta_{j}^{*\top} \boldsymbol x^*_i)^2}{2\sigma_j^2}
\] where \(\boldsymbol x^*_i=(1, \boldsymbol x^\top_i)^\top\) and
\(\boldsymbol\beta^*_{j}=(\beta_{0j}, \bm\beta_j)^\top\).

The relevant derivatives are: \% \[
\frac{\partial\textrm{neglog}\mathcal L^j_{i}}{\boldsymbol \beta^*_{j}}=-\boldsymbol x^*_i\frac{(y_i - \boldsymbol \beta_{j}^{*\top}\boldsymbol x^*_i)}{\sigma_j^2}
\] \[
\frac{\partial\textrm{neglog}\mathcal L^j_{i}}{\sigma_{j}}=\frac{1}{\sigma_j} +
\frac{(y_i - \boldsymbol \beta_{j}^{*\top} \boldsymbol x^*_i)^2}{2}\cdot (-2)\sigma_j^{-3} = 
\frac{1}{\sigma_j} -
(y_i - \boldsymbol\beta_{j}^{*\top} \boldsymbol x^*_i)^2\cdot \sigma_j^{-3}
\]

\subsubsection{Bivariate case}\label{bivariate-case}

\paragraph{Case 1: two ordinal
variables}\label{case-1-two-ordinal-variables}

See derivations for \textbf{mvord}. In principle we have:

\[
\textrm{neglog}\mathcal L^{k,l}_{i} = -\log\left(\Phi_2(U_{i,k}, U_{i,l},\rho) - \Phi_2(U_{i,k}, L_{i,l},\rho) -\Phi_2(U_{i,l}, L_{i,k},\rho)+\Phi_2(L_{i,k}, L_{i,l},\rho) \right)
\] where \[
U_{i,k} = \boldsymbol\psi_k^\top \tilde{\boldsymbol x}^\text{upper}_{i,k},\quad
L_{i,k}  = \boldsymbol\psi_k^\top \tilde{\boldsymbol x}^\text{lower}_{i,k},\quad
U_{i,l}  = \boldsymbol\psi_l^\top \tilde{\boldsymbol x}^\text{upper}_{i,l},\quad
L_{i,l}  =\boldsymbol\psi_l^\top \tilde{\boldsymbol x}^\text{lower}_{i,l}.
\] Let
\(p_i^{k,l} = \Phi_2(U_{i,k} , U_{i,l} ,\rho) - \Phi_2(U_{i,k} , L_{i,l},\rho) -\Phi_2(U_{i,l}, L_{i,k} ,\rho)+\Phi_2(L_{i,k}, L_{i,l} ,\rho)\).

We use here the property that: \[
\frac{\partial\Phi_2(x, y, \rho)}{\partial x} = \phi(x)\Phi\left(\frac{y - \rho x}{\sqrt{1-\rho^2}}\right)
\] and \[
\frac{\partial\Phi_2(x, y, \rho)}{\partial \rho} = \frac{1}{2\pi\sqrt{1-\rho^2}}
\exp\left(\frac{-x^2 - 2\rho xy + y^2}{2(1-\rho^2)}\right).
\]

The relevant derivatives are: \begin{align*}
\frac{\partial\textrm{neglog}\mathcal L^{k,l}_{i}}{\boldsymbol \psi_k}=
  -\frac{1}{p_i^{k,l}}&\left(\left(\frac{\partial\Phi_2(U_{i,k},U_{i,l},\rho)}{\partial U_{i,k}} - \frac{\partial\Phi_2(U_{i,k},L_{i,l},\rho)}{\partial U_{i,k}}\right)\cdot \tilde x^\text{upper}_{i,k} - \right.\\ &\left.\left(\frac{\partial\Phi_2(U_{i,l},L_{i,k},\rho)}{\partial L_{i,k}} - \frac{\partial\Phi_2(L_{i,l},L_{i,k},\rho)}{\partial L_{i,k}}\right)\cdot \tilde x^\text{lower}_{i,k}\right)\\
\end{align*} \% \begin{align*}
\frac{\partial\textrm{neglog}\mathcal L^{k,l}_{i}}{\boldsymbol \psi_l}
  =-\frac{1}{p_i^{k,l}}&\left(\left(\frac{\partial\Phi_2(U_{i,k},U_{i,l},\rho)}{\partial U_{i,l}} - \frac{\partial\Phi_2(U_{i,l},L_{i,k},\rho)}{\partial U_{i,l}}\right)\cdot \tilde x^\text{upper}_{i,l} - \right.\\ &\left.\left(\frac{\partial\Phi_2(U_{i,k},L_{i,l},\rho)}{\partial L_{i,l}} - \frac{\partial\Phi_2(L_{i,l},L_{i,k},\rho)}{\partial L_{i,l}}\right)\cdot \tilde x^\text{lower}_{i,l}\right)\\
\end{align*}

\begin{align*}
\frac{\partial\textrm{neglog}\mathcal L^{k,l}_{i}}{ \rho}&=-\frac{1}{p_i^{k,l}}\left(\frac{\partial\Phi_2(U_{i,k},U_{i,l},\rho)}{\partial \rho} - \frac{\partial\Phi_2(U_{i,k},L_{i,l},\rho)}{\partial \rho} - \frac{\partial\Phi_2(U_{i,l},L_{i,k},\rho)}{\partial \rho} + \frac{\partial\Phi_2(L_{i,l},L_{i,k},\rho)}{\partial \rho}\right)
\end{align*}

\paragraph{Case 2: Bivariate Gaussian}\label{case-2-bivariate-gaussian}

The bivariate negative log likelihood in this case is:

\[
\textrm{neglog}\mathcal L^{k,l}_{i} = \log\left(2\pi\sigma_k\sigma_l\sqrt{1-\rho^2} \right) + \frac{1}{2(1-\rho^2)}\underbrace{\left[\frac{(y_{i,k}-\bm\beta^*_k \bm x^*_i)^2}{\sigma_k^2}-
2\rho\frac{(y_{i,k}-\bm\beta^*_k  \bm x^*_i)(y_{i,l}-\bm\beta^*_l  \bm x^*_i)}{\sigma_k\sigma_l} + 
\frac{(y_{i,l}-\bm\beta^*_l  \bm x^*_i)^2}{\sigma_l^2}
\right]}_{A}
\] The relevant derivatives are: \begin{align*}
\frac{\partial\textrm{neglog}\mathcal L^{k,l}_{i}}{\partial \bm\beta^*_{k}}& =  
\frac{1}{2(1-\rho^2)}\left[\frac{2(y_{i,k}-\bm\beta^*_k \bm x^*_i)}{\sigma_k^2}(-\bm x^*_i)-
2\rho\frac{(y_{i,l}-\bm\beta^*_l \bm x^*_i)}{\sigma_k\sigma_l}(-\bm x^*_i) \right]\\
&=\frac{1}{(1-\rho^2)}(-\bm x^*_i)\left[\frac{(y_{i,k}-\bm\beta^*_k \bm x^*_i)}{\sigma_k^2}-
\rho\frac{(y_{i,l}-\bm\beta^*_l\bm x^*_i)}{\sigma_k\sigma_l} \right]\\
\frac{\partial\textrm{neglog}\mathcal L^{k,l}_{i}}{\partial \bm\beta^*_{l}}& =  
\frac{1}{2(1-\rho^2)}\left[\frac{2(y_{i,l}-\bm\beta^*_l\bm  x^*_i)}{\sigma_l^2}(-\bm x^*_i)-
2\rho\frac{(y_{i,k}-\bm\beta^*_k \bm x^*_i)}{\sigma_k\sigma_l}(-\bm x^*_i) \right]\\
&=\frac{1}{1-\rho^2}(-\bm x^*_i)\left[\frac{(y_{i,l}-\bm\beta^*_l x^*_i)}{\sigma_l^2}-
\rho\frac{(y_{i,k}-\bm\beta^*_k x^*_i)}{\sigma_k\sigma_l} \right]\\
\frac{\partial\textrm{neglog}\mathcal L^{k,l}_{i}}{\partial \sigma_{k}}& =  \frac{1}{\sigma_k} + 
\frac{1}{2(1-\rho^2)}\left[(y_{i,k}-\bm\beta^*_k x^*_i)^2(-2)\sigma_k^{-3}-
2\rho\frac{(y_{i,k}-\bm\beta^*_k \bm x^*_i)(y_{i,l}-\bm\beta^*_l \bm x^*_i)}{\sigma_l}(-\sigma^{-2}_k) \right]\\
&=\frac{1}{\sigma_k} + 
\frac{1}{1-\rho^2}\left[-\frac{(y_{i,k}-\bm\beta^*_k\bm  x^*_i)^2}{\sigma_k^{3}}+
\rho\frac{(y_{i,k}-\bm\beta^*_k \bm x^*_i)(y_{i,l}-\bm\beta^*_l \bm x^*_i)}{\sigma_l\sigma^{2}_k} \right]\\
\frac{\partial\textrm{neglog}\mathcal L^{k,l}_{i}}{\partial \sigma_{l}}& =  \frac{1}{\sigma_l} + 
\frac{1}{2(1-\rho^2)}\left[(y_{i,l}-\bm\beta^*_l\bm  x^*_i)^2(-2)\sigma_l^{-3}-
2\rho\frac{(y_{i,k}-\bm\beta^*_k \bm x^*_i)(y_{i,l}-\bm\beta^*_l \bm x^*_i)}{\sigma_k}(-\sigma^{-2}_l) \right]\\
&=\frac{1}{\sigma_l} + 
\frac{1}{1-\rho^2}\left[-\frac{(y_{i,l}-\bm\beta^*_l x^*_i)^2}{\sigma_l^{3}}+
\rho\frac{(y_{i,k}-\bm\beta^*_k \bm x^*_i)(y_{i,l}-\bm\beta^*_l \bm x^*_i)}{\sigma_k\sigma^{2}_l} \right]\\
\frac{\partial\textrm{neglog}\mathcal L^{k,l}_{i}}{\partial \rho}& =  -\frac{\rho}{1-\rho^2} +\frac{\rho}{(1-\rho^2)^{2}} A - \frac{1}{1-\rho^2}
\frac{(y_{i,k}-\bm\beta^*_k \bm x^*_i)(y_{i,l}-\bm\beta^*_l \bm x^*_i)}{\sigma_k\sigma_l}\\
\end{align*}

\paragraph{Case 3: Mixture of Gaussian and ordinal
responses}\label{case-3-mixture-of-gaussian-and-ordinal-responses}

If we have 1 normal variable at position \(l\) and one ordinal variable
at position \(k\) we have: \[
\textrm{neglog}\mathcal L^{k,l}_{i} = \log\left(\Phi(\eta^u_c) - \Phi(\eta^l_c)\right) + 0.5\log(2\pi)+\log(\sigma_l) + \frac{(y_{il} - \beta_{l}^{*\top} \bm x^*_i)^2}{2\sigma_l^2}
\] where \(\eta^u_c=\frac{\theta_{k, r_{ik}} - \mu_c}{\sigma_c}\) and
\(\eta^l_c=\frac{\theta_{k, r_{ik}-1} - \mu_c}{\sigma_c}\). Let
\(\tilde p_{i}^{k,l} =\Phi(\eta^u_c) - \Phi(\eta^l_c)\).

The relevant derivatives are:

\[
\frac{\partial \textrm{neglog}\mathcal L^{k,l}_{i}}{\partial \bm\psi_k} = - \frac{1}{\tilde p_{i}^{k,l}} \left(\phi(\eta^u_c) \frac{\partial\eta^u_c}{\partial \bm\psi_k} - \phi(\eta^l_c) \frac{\partial\eta^l_c}{\partial \bm\psi_k}\right)
\] Given that \[
\eta^u_c = \frac{\bm\psi_k^\top \tilde{\bm x}^\text{upper}_{i,k} - \frac{\rho}{\sigma_l}(y_{il}-\bm\beta^{*\top}_l\bm x^*_i)}{\sqrt{1-\rho^2}}
\] we have \[
\frac{\partial\eta^u_c}{\partial \bm\psi_k} = \frac{1}{\sqrt{1-\rho^2}}\tilde x^\text{upper}_{i,k}.
\] Similarly, \[
\frac{\partial\eta^l_c}{\partial \bm\psi_k} = \frac{1}{\sqrt{1-\rho^2}}\tilde x^\text{lower}_{i,k}.
\] Putting it all together: \[
\frac{\partial \textrm{neglog}\mathcal L^{k,l}_{i}}{\partial \bm\psi_k} = - \frac{1}{\tilde p_{i}^{k,l}} \left(\phi(\eta^u_c) \frac{1}{\sqrt{1-\rho^2}}\tilde x^\text{upper}_{i,k} - \phi(\eta^l_c) \frac{1}{\sqrt{1-\rho^2}}\tilde x^\text{lower}_{i,k}\right).
\] Now for the coefficients of the normal variable \(\beta^*_l\) \[
\frac{\partial \textrm{neglog}\mathcal L^{k,l}_{i}}{\partial \bm\beta^*_l} = - \frac{1}{\tilde p_{i}^{k,l}}\left(\phi(\eta^u_c) \frac{\partial\eta^u_c}{\partial \bm\beta^*_l} - \phi(\eta^l_c) \frac{\partial\eta^l_c}{\partial \bm\beta^*_l}\right) -\bm x^*_i
\frac{(y_{il} - \beta^{*\top}_l \bm x^*_i)}{\sigma_l^2}
\] where we have \[
\frac{\partial\eta^u_c}{\partial \bm\beta^*_l} = \frac{\partial\eta^l_c}{\partial \bm\beta^*_l} =
\frac{\rho}{\sigma_l\sqrt{1-\rho^2}} \bm x^*_i.
\]

Putting it all together:

\[
\frac{\partial \textrm{neglog}\mathcal L^{k,l}_{i}}{\partial \bm\beta^*_l} = - \frac{1}{\tilde p_{i}^{k,l}}\frac{\rho}{\sigma_l\sqrt{1-\rho^2}}x^*_i\left(\phi(\eta^u_c)  - \phi(\eta^l_c) \right) -x^*_i
\frac{(y_{il} -  \bm\beta^{*\top}_l  \bm x^*_i)}{\sigma_l^2}
\] For the correlation parameter of the pair: \[
\frac{\partial \textrm{neglog}\mathcal L^{k,l}_{i}}{\partial \rho} = -\frac{1}{\tilde p_{i}^{k,l}} \left(\phi(\eta^u_c) \frac{\partial\eta^u_c}{\partial \rho} -\phi(\eta^l_c)\frac{\partial\eta^l_c}{\partial \rho}\right)
\] where let
\(g^u(\rho)=\psi_k^\top \tilde x^\text{upper}_i - \frac{\rho}{\sigma_l}(y_{il}-\beta^{*\top}_lx^*_i)\)
and \(h(\rho)=\sqrt{1-\rho^2}\),
\(\partial h(\rho)/\partial \rho = - \rho/\sqrt{1-\rho^2}\).

\[
\frac{\partial\eta^u_c}{\partial \rho} =\frac{-(\sqrt{1-\rho^2})
 (y_{il} - \beta_l^{*\top} x^*_i)/\sigma_l + \rho/\sqrt{1-\rho^2}\cdot g^u(\rho)}{1-\rho^2}=-\frac{(y_{il} -  \bm\beta_l^{*\top}  \bm x^*_i)/\sigma_l}{\sqrt{1-\rho^2}} + \frac{\rho \eta^u_c}{1-\rho^2}
\] and \[
\frac{\partial\eta^l_c}{\partial \rho} =-\frac{(y_{il} -  \bm\beta_l^{*\top}  \bm x^*_i)/\sigma_l}{\sqrt{1-\rho^2}} + \frac{\rho \eta^l_c}{1-\rho^2}
\]

Finally, for the standard deviation parameter of the normal: \[
\frac{\partial \textrm{neglog}\mathcal L^{k,l}_{i}}{\partial \sigma_l} = -\frac{1}{\tilde p_{i}^{k,l}} \left(\phi(\eta^u_c) \frac{\partial\eta^u_c}{\partial \sigma_l} -\phi(\eta^l_c)\frac{\partial\eta^l_c}{\partial \sigma_l}\right) +
\frac{1}{\sigma_l} -
(y_{il} -  \bm\beta_{l}^{*\top}  \bm x^*_i)^2\cdot \sigma_l^{-3}
\]

\[
\frac{\partial\eta^u_c}{\partial\sigma_l} =\frac{\partial\eta^l_c}{\partial\sigma_l} 
=\frac{\rho (y_{il} -  \bm\beta_{l}^{*\top} \bm  x^*_i)}{\sigma_c\sigma_l^{2}}
\]

\end{document}